\documentclass[prd,footinbib,twocolumn,floatfix]{revtex4}
\usepackage{amsmath}
\usepackage{amssymb}
\usepackage{graphics}

\makeatletter
\renewcommand{\@makecaption}[2]{% #1 is Figure Num, #2 is caption
  \vskip\abovecaptionskip
  \sbox\@tempboxa{\small\sf #1: #2}%
  \ifdim \wd\@tempboxa >\hsize
  \small\sf #1: #2\par
  \else
    \global \@minipagefalse
    \hb@xt@\hsize{\hfil\box\@tempboxa\hfil}%
  \fi
  \vskip\belowcaptionskip}
\makeatother

\def\ba{\begin{eqnarray}}
\def\ea{\end{eqnarray}}

\def\hat{\widehat}
\def\tilde{\widetilde}

\def\Dslash{\,\,{\raise.15ex\hbox{/}\mkern-12mu D}}
\def\Dbarslash{\,\,{\raise.15ex\hbox{/}\mkern-12mu {\bar D}}}
\def\delslash{\,\,{\raise.15ex\hbox{/}\mkern-9mu \partial}}
\def\delbarslash{\,\,{\raise.15ex\hbox{/}\mkern-9mu {\bar\partial}}}
\def\pslash{\,\,{\raise.15ex\hbox{/}\mkern-9mu p}}
\def\calDslash{\,\,{\raise.15ex\hbox{/}\mkern-12mu {\cal D}}}

\newcommand\im{{\rm Im}}
\newcommand{\ZZ}{{\mathbb Z}}

\newcommand{\RR}{{\mathbb  R}}
\newcommand{\hF}{{\hat F}}
\newcommand{\hA}{{\hat A}}

\newcommand{\hG}{{\hat G}}
\newcommand{\ra}{\rightarrow}
\newcommand{\hH}{{\hat H}}
\newcommand{\hatt}{{\hat t}}
\newcommand{\hm}{{\mathfrak m}}
\newcommand{\hB}{{\hat B}}

\newcommand{\dt}{{\underline{t}}}
\newcommand{\da}{{\underline{\alpha}}}
\newcommand{\frg}{{\mathfrak g}}
\newcommand{\frh}{{\mathfrak h}}
\newcommand{\Hom}{{\rm Hom}}
\newcommand{\tg}{{\tilde g}}
\newcommand{\tG}{{\tilde G}}

\renewcommand{\im}{{\rm im}}
\renewcommand{\ker}{{\rm ker}}
\newcommand{\hf}{{\mathfrak f}}
\newcommand{\Aut}{{\rm Aut}}

\begin{document}
%\preprint{CALT-XX-XXXX}

\title{Topological Quantum Field Theory, Nonlocal Operators, and Gapped Phases of Gauge Theories}

\author{Sergei Gukov, Anton Kapustin}

\affiliation{California Institute of Technology, Pasadena, CA 91125, USA}

%\author{Nathan Seiberg}
%\email{seiberg@ias.edu}
%\affiliation{School of Natural Sciences, Institute for Advanced Study, Princeton, NJ 08540, USA}

%\pacs{}
%\keywords{}

\begin{abstract}
We revisit the role of loop and surface operators as order parameters for gapped phases of four-dimensional gauge theories.  We show that in some cases surface operators are confined, and that this fact can be used to distinguish phases which are not distinguished by the Wilson-'t Hooft criterion. The long-distance behavior of loop and surface operators which are neither confined nor screened is controlled by a 4d TQFT. We construct these TQFTs for phases which are characterized by the presence of electrically and/or magnetically charged condensates. Interestingly, the TQFT describing a phase with a nonabelian monopole condensate is based on the theory of nonabelian gerbes. We also show that in phases with a dyonic condensate the low-energy theta-angle is quantized.
\end{abstract}

\maketitle

\newcommand{\be}{\begin{equation}}
\newcommand{\ee}{\end{equation}}

\section{Introduction}

Non-local operators can be excellent order parameters distinguishing between gapped phases of four-dimensional field theories.
Familiar examples include loop operators, {\it i.e.} non-local operators supported on
one-dimensional submanifolds in space-time.
For example, 't Hooft loop operators can detect spontaneous symmetry breaking
and exhibit an area law in Higgs phases. Likewise, in a confining phase a QCD string
can end on a Wilson loop operator which, therefore, can exhibit an area law and serves as an order parameter
for confinement. One can take the area law for Wilson (resp.  't Hooft) loop operators as a definition of a confining (resp. Higgs) phase. We will refer to this as the Wilson-
't Hooft classification of phases. Yet, there exist confining and Higgs phases which can not be distinguished
by the standard loop operators~\cite{Witten3dindex,CSW}.
In situations like this, one might expect that other non-local operators, such as surface operators, come to the rescue. 

Another approach to the classification of gapped phases is based on Topological Quantum Field Theory (TQFT). One expects that at long distances any gapped phase is described by a TQFT, and the isomorphism class of this TQFT can be used to label this gapped phase. Nonlocal observables in such a TQFT are those gauge theory observables which are neither confined nor screened (i.e. they have long-range effects on other observables.) This approach to the classification seems complementary to the Wilson-'t Hooft approach, but in fact incorporates it and is more powerful. For example, we will see that when a TQFT is formulated in terms of the same gauge fields as the UV theory, the loop operators which exhibit an area law in the full theory fail to be gauge-invariant with respect to some additional gauge symmetries present in the TQFT. This is a TQFT manifestation of the area law. In particular, the TQFT description is able to capture the information about the confinement index introduced in \cite{CSW}. But the TQFT description provides more information; for example, it allows one to compute the ground-state degeneracy on an arbitrary 3-manifold, and also encodes the Aharonov-Bohm phases of the nonlocal operators which are neither confined nor screened. Thus it is of interest to try to classify 4d TQFTs involving gauge fields and identify which gapped phases they correspond to. This problem is not as hopeless as one might think since one is interested only in unitary TQFTs, and such theories are scarce in space-time dimension higher than three.

In this paper we take some modest steps in this direction. We start our discussion with surface operators in abelian gauge theories and how their behavior helps to distinguish gapped phases which look the same from the point of view of the Wilson-'t Hooft criterion. Then we show how to formulate TQFTs which describe abelian gapped phases with arbitrary dyonic condensates.  Interestingly, the theta-angle is quantized in these theories.  We also write down a TQFT  which describes massive phases where the microscopic  abelian gauge group is partially confined and partially Higgsed; the TQFT contains information about the confinement index. Turning to nonabelian gauge theories, we write down TQFT actions which describe arbitrary Higgs phases.  Finally we propose a TQFT which describe confining phases of nonabelian gauge theories. Surprisingly, they involve nonabelian  gerbes, but appear to be equivalent to the Higgs TQFTs for abelian unbroken gauge groups.  The theta-angle is again quantized. We conclude by discussing the implications of our work for electric-magnetic duality and the realization of various phases in lattice gauge theory. In particular, we interpret various passible lattice formulations of $SU(N)/\ZZ_N$ gauge theories as coupling a lattice $SU(N)$ gauge theory to a TQFT. Related remarks appear in section 6 of Ref.~\cite{AST}.

While in this paper we limit ourselves to phases with a mass gap, the results may have application to Coulomb phases as well (i.e. to phases where the only massless particles are photons and perhaps their superpartners), like those considered in \cite{CSW}. Indeed, if one takes the gauge coupling for the photons to zero, one can treat the abelian gauge field as non-dynamical, and therefore describe the low-energy theory as a TQFT with an abelian global symmetry. The only non-uniqueness in this procedure is related to the existence of the electric-magnetic duality in the low-energy theory and the attendant ambiguity in the definition of the low-energy coupling.

\section{Surface operators in abelian gauge theories}

In a theory with gauge group $G=U(1)$ there are two basic surface operators:
``electric'' surface operators (a.k.a. surface operators of type $\eta$)
that, much like Wilson operators, can be described by inserting  into the path integral
a function of fundamental degrees of freedom (fields) in the theory~\cite{GW,RTN}:
\be
\exp \left( i \eta \int_{\Sigma} F \right) \,,
\label{etaangle}
\ee
and ``magnetic'' surface operators (a.k.a. a surface operator of type $\alpha$)
which, like 't Hooft operators, can be defined by requiring the gauge field $A$ (and, possibly, other fields)
to have a prescribed singularity along a surface~$\Sigma$:
\be
F = 2 \pi \alpha \delta^2_{\Sigma} +{\rm smooth}\,.
\label{asing}
\ee
Here, $F$ is the curvature 2-form and $\delta_{\Sigma}$ is a 2-form delta function
that is Poincar\'e dual to $\Sigma$. The continuous parameters $\alpha$ and $\eta$ are both periodic, {\it i.e.} $U(1)$-valued. More precisely, $\alpha$ is defined up to elements of the cocharacter lattice and $\eta$ is defined up to elements of the character lattice of $G$. Both lattices will play an important role in the discussion below and simply are copies of $\ZZ$ in the present example.

Let us consider a Higgs phase where the $U(1)$ gauge group is spontaneously broken down to $\ZZ_n$. This happens {\it e.g.} in a simple model with a complex scalar  field $\phi$ of charge $n$ which acquires a vacuum expectation value. In such a theory there are Abrikosov-Nielsen-Olesen (ANO) flux tubes whose magnetic flux is quantized in units of $2\pi/n$. If one inserts a magnetic monopole of charge $p$ into this phase, its flux is collimated into an ANO flux tube whose charge is $p n$ times the flux quantum. The corresponding 't Hooft loop will obey an area law, with the tension given by the tension of the ANO flux tube. All Wilson loops, on the other hand, will obey the perimeter law. Thus from the point of view of the Wilson-'t Hooft classification all such phases are the same, regardless of the value of $n$.

To detect $n$, we may consider surface operators known in this context as Alice strings \cite{ASSchwarz, BGHHS, WilczekKrauss, PreskillKrauss}. A magnetic surface operator of type $\alpha$ is a singular magnetic flux tube with a magnetic flux $2\pi\alpha$. Consider inserting such a flux tube along a straight line $\ell\subset\RR^3$. In order to have finite energy, the covariant derivative $D\phi=(d-i e n A)\phi$ must rapidly approach zero far from $\ell$. This is only possible if $n\alpha\in\ZZ$. Even if this condition is not satisfied, $|D\phi|$ will decay as $1/r$ only, resulting in a logarithmically divergent energy. If $\ell$ is replaced by a closed contour of size $L$, the energy will be finite and scale like $\log L$. Thus magnetic surface operators with $n\alpha\notin\ZZ$ will be logarithmically confined. Note that there is no linear confinement in this model, i.e. no volume law for magnetic surface operators.
On the other hand, electric surface operators do not create any long-range distortions in the condensate and are screened (and obey the area law). 

Magnetic surface operators with $\alpha=k/n$, $k=1,\ldots, n-1$, are not confined and can be used to measure the conserved $\ZZ_n$-valued charge carried by Wilson loops. This charge is what remains of the $\ZZ$-valued charge of the UV theory after one takes into account screening due to the condensate.  Magnetic surface operators are worldsheets of what is known as Alice strings.

By electric-magnetic duality, we expect that when monopoles of charge $n$ condense, electric surface operators of type $\eta$ are logarithmically confined unless $n\eta\in\ZZ$. In such a phase all Wilson loops obey an area law, regardless of their charge, while all 't Hooft loops obey a perimeter law. But while 't Hooft loops with magnetic charge divisible by $n$ are screened, magnetic charge modulo $n$ can be detected at long distances by unconfined electric surface operators. Thus in such a phase electric surface operators with $\eta=k/n$, $k=1,\ldots,n-1$ are Alice strings.

\section{TQFTs for abelian gapped phases}

\subsection{Abelian Higgs phases}

Consider again the model with gauge group $U(1)$ where a complex scalar of charge $n$ condenses. As discussed above, nonlocal operators surviving in the long-distance limit are Wilson loops with a $\ZZ_n$-valued charge and magnetic surface operators with $\alpha=k/n$, $k=0,1,\ldots,n-1$ (of course, the surface operator with $\alpha=0$ is trivial). Their correlators are encoded by a TQFT, namely gauge theory with gauge group $\ZZ_n$. This TQFT has several equivalent continuum descriptions \cite{MSM,TomNati} (see also \cite{HellermanES} for a related perspective). 

One possibility is to use an action
\be
S=\int (d\varphi-n A) \wedge h,
\ee
where $\varphi$ is the phase of the complex scalar, $A$ is the $U(1)$ gauge field, and $h$ is a Lagrange multiplier 3-form. The fields have the following gauge transformations:
\be
A\mapsto A+df,\quad \varphi\mapsto \varphi+n f,
\label{agauge}
\ee
where $f$ is a function with values in $\RR/2\pi\ZZ$.
This action enforces the condition $D\varphi=0$ everywhere. This condition forces $F=0$ and therefore one cannot have any 't Hooft loops in this TQFT. This is how the TQFT knows about the area law for 't Hooft loops. The same condition also ensures that electric surface operators are trivial. On the other hand, magnetic surface operators make sense only if $n\alpha\in\ZZ$. This is how TQFT knows about the logarithmic confinement of all other magnetic surface operators. Finally, a Wilson loop whose charge $q$ is divisible by $n$ can be written as 
\be
\exp\left(\frac{i q}{n}\int d\varphi\right)=1,
\ee
and thus is trivial. 

Another possibility is to dualize $\varphi$ into a 2-form $B$, so that the action reads
\be
S=\frac{n}{2\pi} \int B\wedge dA.
\ee
This action describes the same physics, but from a slightly different perspective. Namely, a Wilson loop $\ell$ whose charge $q$ is divisible by $n$ is now trivial because it can be eliminated by a 1-form gauge transformation $B\mapsto B+d\lambda$, where $\lambda$ is a connection 1-form on a $U(1)$ bundle which corresponds to an 't Hooft loop insertion along $\ell$. Note, the gauge transformation for $B$ is completely independent of (\ref{agauge}). On the other hand, magnetic surface operators which do not satisfy $n\alpha\in\ZZ$ are not invariant under 1-form gauge transformations: in the presence of a surface operator supported on a submanifold $\Sigma$  the integrand in the path-integral is multiplied by a factor
\be
\exp\left(i n \alpha \int_\Sigma d\lambda \right),
\ee
which is equal to $1$ for all conceivable $\Sigma$ and $\lambda$ only if $n\alpha\in\ZZ$. 

It is easy to generalize this to arbitrary (connected) abelian gauge groups and arbitrary condensates. Such a group is a torus $G\simeq U(1)^N$. Charges of fields which have condensed define a sub-lattice $\Gamma_0$ in the charge lattice $\Gamma=\Hom(G, U(1))\simeq H^1(G,\ZZ)$. If we assume that all the generators of the gauge group are broken (i.e. the phase is a Higgs phase rather than a mixed Higgs-Coulomb phase), then $\Gamma_0$ must be a finite-index subgroup in $\Gamma$. Conserved electric charge takes values in the finite abelian group $\Gamma/\Gamma_0$, and the TQFT describing the long-distance behavior is a gauge theory with a finite abelian gauge group $G_0=\Hom(\Gamma/\Gamma_0, U(1))$. Such a theory can be described in the continuum as follows. As usual, in order to break $G$ down to $G_0$, one needs to introduce Higgs fields valued in $H=G/G_0$. In our case, $G$ is a torus, and $G_0$ is its finite subgroup, so $H$ is also a torus.
The action is
\be
S=\int \langle d\varphi - s(A) , \wedge h \rangle \,,
\ee
where $s$ is the projection map from $G$ to $H$ and $h$ is a 3-form with values in the dual of the Lie algebra of $H$. Magnetic surface operators in the UV theory are labeled by $\alpha\in G$; such an operator is confined unless $s(\alpha)=0$ (i.e. unless $s(\alpha)$ is the identity element in $H$). Thus in the TQFT magnetic surface operators are labeled by elements of the finite abelian group $\ker (s)=G_0$. The conserved charge for Wilson loops takes values in the finite abelian group $\Gamma/\Gamma_0$ which is Pontryagin dual to $G_0$. 

As before, we can dualize periodic scalars $\varphi$ taking values in $H$ into a connection 2-form $B$ with values in the dual of the Lie algebra of $H$. The action is
\begin{equation}\label{BF}
S=\frac{i }{2\pi} \int \langle B , \wedge s(F) \rangle \,.
\end{equation}
Gauge transformations of $A$ and $B$ are
$$
A\mapsto A+df,\quad B\mapsto B+d\lambda,
$$
where $f$ is an arbitrary periodic scalar with values in the torus $G$, and $\lambda$ is an arbitrary abelian gauge field with a gauge group $\hH=\Hom(H,U(1))$. The corresponding Chern class $d\lambda/2\pi$ takes values in the charge lattice of $H$ (or equivalently, in $H_1(\hH,\ZZ)$). In the presence of a magnetic surface operator  labeled by $\alpha\in G$ and inserted along a submanifold $\Sigma$, such a gauge transformation multiplies the integrand in the path-integral by a factor
\be
\exp\left(i \langle s(\alpha) , \int_\Sigma d\lambda\rangle \right).
\ee
This is trivial for all $\Sigma$ and $\lambda$ if and only if $\alpha\in\ker\, s$. Thus magnetic surface operators are labeled by elements of $\ker (s)=G_0$. 

't Hooft loops in such a theory are not invariant under 1-form gauge transformations, which is how the TQFT represents their confinement. Wilson loops whose charge lies in the subgroup $\Gamma_0$ can be eliminated by 1-form gauge transformations, which means that these Wilson loops are screened. Thus the conserved electric charge takes values in the quotient group $\Gamma/\Gamma_0$ which is Pontryagin-dual to $\ker\, s$.

\subsection{Abelian confining phases}

To obtain TQFTs describing confining phases of abelian gauge theories, we can now dualize the gauge field $A$ in the action (\ref{BF}). The resulting TQFT has an action
\be
S=\int \langle \hF-\hatt (B) , \wedge b \rangle \,,
\ee
where $\hF$ is the curvature 2-form of a dual gauge field $\hat A$ with gauge group $\hG=\Hom(G,U(1))$. The 2-form gauge field $B$ is the same as before, i.e. it is associated with 1-form gauge transformations which are parameterized by an abelian gauge field with gauge group $\hH=\Hom(H,U(1))$. The map $\hatt: \hH\ra\hG$ is the dual of $s:G\ra H$. The gauge field $\hA$ now also transforms under the 1-form gauge transformations:
$$
\hA\mapsto \hA+\hatt(\lambda),
$$
so that the action is gauge-invariant.

Note that while $G$ covers $H$, with $s$ being the covering map, after dualization the situation is reversed, {\it i.e.} $\hH$ covers $\hG$, with $\hatt$ being the covering map. We conclude therefore that confining phases of an abelian gauge theory with the UV gauge group $\hG$ are classified by finite covers $\hatt: \hH\ra \hG$. 

Let us see how the physics of monopole condensation is encoded in this datum. First consider magnetic charges. In the UV theory magnetic charges take values in $\pi_1(\hG)=H_1(\hG,\ZZ)$. The map $\hatt: H_1(\hH,\ZZ)\ra H_1(\hG,\ZZ)$ is injective, and its image is a finite index subgroup of $H_1(\hG,\ZZ)$. This subgroup can be identified with the charges of monopoles which have condensed. Indeed, since the gauge feld $\hA$ transforms under 1-form gauge transformations, the 't Hooft flux is now defined only modulo elements of $\im\, \hatt$. Consequently, we expect that 't Hooft loops whose charge lies in $\im\, \hatt$ are screened. Indeed, consider an insertion of an $H$-monopole at a point $p$ in space-time. By definition, this is a point where the B-field is singular in such a way that $\int_V dB=2\pi \hm$ where $V\simeq S^3$ is the boundary of a small ball centered at $p$ and $\hm\in H_1(\hH,\ZZ)$ is the $H$-monopole charge. In the neighborhood of such a point the field $\hB$ cannot be regarded as a smooth 2-form: there should exist points on $V$ such that the flux of $B$ on small 2-spheres around these points adds up to $2\pi \hm$. The equation of motion $\hF=\hatt (B)$ implies that the same is true for the 2-form $\hF$, except that the flux of $\hF$ must add up to $\hatt (\hm)$. In particular, this means that an 't Hooft loop operator may terminate at an $H$-monopole if its magnetic charge is equal to $\hatt (\hm)$ for some $\hm\in H_1(\hH,\ZZ)$. Thus 't Hooft flux is conserved only modulo elements of $\im\, \hatt$. In other words, conserved 't Hooft flux takes values in the finite abelian group $H_1(\hG,\ZZ)/H_1(\hH,\ZZ)$. 

Second, none of the Wilson loops are invariant under 1-form gauge transformations, which is how the TQFT accounts for the confinement of electric flux in the UV theory. Third, an electric surface operator in the UV theory is labeled by $\eta$ which takes values in $\Hom(\hG,U(1))\simeq G$. It is invariant under 1-form gauge transformations if and only if the sublattice $\im\, \hatt\subset H_1(\hG,\ZZ)$ annihilates $\eta$. This is a reflection of the fact that all other electric surface operators are confined. Thus the charge of a gauge-invariant electric surface operator takes values in a finite abelian subgroup of $\Hom(\hG,U(1))$ which is Pontryagin dual to the group where conserved 't Hooft flux takes values.

\subsection{Oblique confining phase}

A phase with a dyon condensate is called an oblique confinement phase.  Here by a dyon we mean a particle which carries both electric and magnetic charge. In the case when the UV gauge group is $U(1)$, we can guess the TQFT for such a phase by making use of the Witten effect \cite{Witteneffect}. Namely, since in the presence of a theta-angle a monopole carries both magnetic and electric charges, it is natural to consider the following action:
\be
S=\int \langle \hF-n B , \wedge b \rangle +\frac{i m}{4\pi n}\int \text{Tr} \hF \wedge \hF
\ee
The parameter $m$ must be integer to ensure the invariance of $\exp(-S)$ under 1-form gauge transformations
$$
B\mapsto B+d\lambda,\quad \hA \mapsto \hA + n\lambda.
$$
To see that this action describes a phase with a dyon condensate of charge $(m,n)$, consider how the action transforms under 1-form gauge transformations. The first terms is obviously invariant. If $\hF$ is everywhere nonsingular, then using $d\hF=0$ and integration by parts, we see that the last term is also invariant. However, in the presence of an 't Hooft loop operator inserted along $\ell$ the theta-term fails to be gauge-invariant. Rather, the action is shifted by
\be
i mq\int_\ell \lambda,
\ee
where $q$ is the magnetic charge of an 't Hooft loop. To make a gauge-invariant loop operator, one need to multiply an 't Hooft loop with charge $q$ by a Wilson loop with charge $p$ such that $mq=np$. Therefore the most general gauge-invariant loop operator allowed in this model is a Wilson-'t Hooft loop with charges $(p,q)$ such that $p=k m/gcd(m,n),$ $q=kn/gcd(m,n)$, and $k\in\ZZ$. Such a loop operator can terminate on an $H$-monopole of charge $\hm$ if and only if $q=n \cdot \hm$ and accordingly $p=m\cdot \hm$. We interpret this as a presence of a condensate of dyons with charges $(m,n)$. Consequently, a Wilson-'t Hooft loop operator with $p=k m/gcd(m,n),$ $q=kn/gcd(m,n)$, is screened if $k$ is divisible by $gcd(m,n)$. Only the value of $k$ modulo $gcd(m,n)$ can be regarded as a conserved charge.

A general surface operator in a $U(1)$ gauge theory depends on two parameters, $\alpha$ and $\eta$, both taking values in $\RR/\ZZ$ \cite{GW}. The parameter $\alpha$ determines the singularity in the gauge field  (\ref{asing}), while $\eta$-dependence enters through the factor (\ref{etaangle}). 1-form gauge-invariance puts a constraint on these two parameters:
\begin{equation}\label{etaalphagaugeinv}
\exp(2\pi i(\eta n-m\alpha))=1.
\end{equation}
1-form gauge transformations also lead to additional identifications on the parameter space $\RR/\ZZ\times\RR/\ZZ$:
$$
\alpha\mapsto \alpha+n a,\quad \eta\mapsto \eta+m a,\quad a\in\RR.
$$
We can partially ``fix a gauge'' by setting $\alpha=0$. Then the condition (\ref{etaalphagaugeinv}) reduces to $\exp(2\pi i n\eta)=1$. We have a residual $\ZZ_n$ gauge-invariance, with $a=p/n$, $p\in\ZZ/n\ZZ$, which acts on $\eta$ as follows:
\be
\eta\ra\eta+pm/n.
\ee
Modding out by this symmetry, we get $gcd(m,n)$ inequivalent values for $\eta$. Thus $\eta$ can be thought as taking values in $\ZZ_{gcd(m,n)}$. This result was to be expected, as surface and loop operators should be labeled by elements of Pontryagin dual groups. 

\subsection{Mixed phases}

More generally, we may consider an abelian gauge group $G\simeq U(1)^N$ and a general condensate of electrically and magnetically charged fields. In this case one must account for a possibility that the gauge group is partially confined and partially Higgsed. This happens if the condensate magnetic charges do not span a finite index subgroup in $H_1(G,\ZZ)$. Then some of the gauge fields are invariant under 1-form gauge transformations and have to be Higgsed if one is to get a gapped phase. The most general action which accomplishes this involves both 2-form gauge fields $B$ taking values in the tangent space to a torus $H$ and periodic scalars $\varphi$ taking values in a torus $\Phi$ and has the form
\begin{eqnarray}
S & = & \int \langle F-t(B) , \wedge b \rangle
+ \int \langle d\varphi-s(A) , \wedge h \rangle \nonumber \\
& & +\frac{i}{8\pi^2}\int \theta(F,\wedge F)+i \int \psi(d\phi, \wedge  dB) \,.
\end{eqnarray}
Here $t:H\ra G$ and $s:G\ra \Phi$ are homomorphisms of abelian groups and $\theta$ is a bilinear form on the Lie algebra of $G$. The last term can be eliminated by shifting the Lagrange multiplier fields $h$ and $b$, so we may set $\psi=0$. One can assume that the map $s$ is surjective, since those $\Phi$ which are not in the image of $s$ can be trivially integrated out. Gauge transformations take the form
$$
B\mapsto B+d\lambda,\quad A\mapsto A+t (\lambda)+d f,\quad \varphi\mapsto\varphi+s(f),
$$
where $\lambda$ is a gauge field with gauge group $H$ and $f$ is a function with values in $G$. 
Gauge-invariance of the action requires $s\circ t=0$ and also imposes a quantization condition on the bilinear form $\theta$:
$$
\theta(t(\hm),\hf)\in 2\pi\ZZ,\quad \forall \hf\in H_1(G,\ZZ), \quad \forall \hm\in H_1(H,\ZZ)
$$

The meaning of the data $t,s,$ and $\theta$ is clear: they describe the electric and magnetic charges of the particles in the condensate. 

Let us analyze the observables for this TQFT. For simplicity, let us assume that $\theta$-angles vanish. The Higgs fields $\Phi$ break the gauge group down to a subgroup $\ker\, s$. This group is abelian but not necessarily connected. The gauge fields for the subgroup $\im\, t$ transform non-trivially under 1-form gauge transformation which we interpret as confinement. Since $s\circ t=0$, the confined subgroup is contained in the unbroken subgroup $\ker\, s$. Thus the part of the gauge group which is neither Higgsed nor confined is a sub-quotient $\ker\, s/\im\, t$. Accordingly, unconfined Wilson loops are labeled by characters of $\ker\, s/\im\, t$. 

This model exhibits a non-trivial confinement index, in general, in the sense that several confined Wilson loops can combine into a Wilson loop which is not confined. Indeed, a
a Wilson loop with a microscopic charge $q\in \Hom(G,U(1))$ is confined if and only if it is invariant under 1-form gauge transformations, i.e. if $q\circ t\neq 0$. Clearly, it may well happen that $q_1\circ t\neq 0$ and $q_2\circ t\neq 0$, but $(q_1+q_2)\circ t=0$. 

There are also 't Hooft loops; as before, they are labeled by element of $\ker\, t$. It is easy to see that there are also magnetic and electric surface operators labeled by elements of the Pontryagin-dual groups.

\section{TQFTs for nonabelian gapped phases}

\subsection{Nonabelian Higgs phases}

Consider a gauge theory with a nonabelian gauge group $G$ which is Higgsed down to a finite subgroup $G_0$. The gapped phase is described by TQFT which is a gauge theory with gauge group $G_0$. A continuum description can obtained by generalizing the abelian action. The analog of $\varphi$ takes values in the manifold $G/G_0$. This manifold is a group only if $G_0\subset G$ is a normal subgroup, which need not be the case. The TQFT action is
$$
S=\int \langle D\varphi, h\rangle,
$$
where $D \varphi$ is the covariant derivative of $\varphi$ regarded as a section of a $G$-bundle $\mathcal E$ with fiber $G/G_0$, and $h$ is a 3-form with values in the dual of the vertical tangent bundle of $\mathcal E$. The equation of motion $D\varphi=0$ says that $\varphi$ is a trivialization of $\mathcal E$, which means that on-shell the structure group is reduced to $G_0$. 

't Hooft operators in this theory violate equations of motion, which is how the TQFT represents their confinement. Wilson loops can be defined and are labeled by representations of $G$, but since the holonomy of the connection is forced to lie in $G_0$, only the decomposition of a representation with respect to $G_0$ plays a role. In other words, Wilson loops corresponding to representations of $G$ on which $G_0$ acts trivially are screened. 

Magnetic surface operators are labeled by conjugacy classes of elements of $G_0$. As for electric surface operators, they are all trivial, since the gauge field $A$ is forced to be flat by the equations of motion.

\subsection{Nonabelian confining phases}

\subsubsection{Monopole condensation}

If the UV gauge group $G$ is nonabelian, one cannot get a TQFT action describing a confining phase by dualizing a TQFT action for a Higgs phase. Instead we propose an action based on the idea that confinement is associated with monopole condensation. 

Monopoles have topological charge ('t Hooft flux) taking values in $\pi_1(G)$, where $G$ is the high-energy gauge group\footnote{We are assuming that $G$ is connected, since this is by far the most common case. It would be interesting to extend the discussion that follows to disconnected groups, such as $O(N)$.}. Condensed monopoles define a subgroup $\Gamma_0$ of $\pi_1(G)$, and in the presence of such a condensate one should only be able to define 't Hooft flux modulo the elements of $\Gamma_0$. Geometrically, a choice of a subgroup $\Gamma_0\subset\pi_1(G)$ corresponds to a covering homomorphism $t:H\ra G$ with fiber $\ker\, t=\pi_1(G)/\Gamma_0=\pi_1(G)/\pi_1(H)$. Well-defined t' Hooft fluxes in the low-energy theory should be labeled precisely by elements of $\ker\, t$. Note that although $G$ may be nonabelian, $\pi_1(G)$ and therefore $\Gamma_0$ and $\ker\, t$ are necessarily abelian.

Clearly, $\ker\, t$ can be infinite only if $G$ contains $U(1)$ factors. If $\ker\, t$ is infinite, 't Hooft flux in some of these $U(1)$ factors is well-defined.  This means that the monopoles in the condensate are not charged with respect to these $U(1)$ factors, and consequently these $U(1)$ factors are not confined. Thus it is natural to restrict to the case where $\ker\, t$ is finite. To summarize,  we expect that confining phases of a gauge theory with high-energy gauge group $G$ are classified by finite covers $t:H\ra G$.

The simplest possibility is $H=G$, with the identity homomorphism as the covering map. If $G$ is a simple group, one can also take $H$ to be its universal cover. If $G=SO(3)$, these are the only possibilities, i.e. $H=SO(3)$ or $H=SU(2)$. The first possibility corresponds to the condensation of monopoles with all possible 't Hooft fluxes, so in the low-energy phase one cannot define a conserved 't Hooft flux at all. The second possibility corresponds to the situation when condensing monopoles have trivial 't Hooft flux (but nonzero GNO flux) , so in the low-energy phase one can define $\pi_1(G)$-valued 't Hooft flux. 

In a confining phase there are electric surface operators which measure conserved 't Hooft flux. Such operators are labeled by elements of the group $\hat\Gamma=\Hom(\ker\, t, U(1))$, i.e. the Pontryagin-dual of $\Gamma$. Note that in the high-energy theory analogous surface operators are labeled by elements of $\widehat{\pi_1(G)}=\Hom(\pi_1(G),U(1))$. One can describe $\hat\Gamma$ as a subgroup of $\widehat{\pi_1(G)}$ which consists of surface operators which are trivial (equal to $1$) on the subgroup $\Gamma_0=\pi_1(H)$. That is, in the low-energy theory one has only those surface operators which do not detect the fluxes of condensed monopoles.

\subsubsection{Nonabelian gerbes and nonabelian B-fields}

A suitable topological field theory for confining phases can be constructed using the theory of nonabelian B-fields which ``categorifies'' the theory of connections on principal bundles (see \cite{BaezSchreiber} for a review). The higher analog of a Lie group  is known as a Lie 2-group. This is a quadruple $(G,H,t,\alpha)$ where  $G$ and $H$ are Lie groups, $t:H\ra G$ is a homomorphism, and $\alpha:G\ra \Aut(H)$ is another homomorphism. Here $\Aut( H)$ is the group of automorphisms of $H$. These data should satisfy the following compatibility conditions:
$$
t(\alpha(g)(h))=gt(h)g^{-1},\quad hh'h^{-1}=\alpha(t(h))(h') %,\quad\forall h,h'\in H,\quad\forall g\in G.
$$
for all $h,h'\in H$ and for all $g \in G$.
It follows from these conditions that $\ker\, t$ always lies in the center of $H$. 

The most obvious Lie 2-group  is the one associated to a central extension of $G$ by an abelian group, or equivalently to a surjective homomorphism $t:H\ra G$ such that $\ker\, t$ is central. This is precisely the datum which describes a choice of a monopole condensate for a gauge group $G$, as discussed above. We would like to argue that to such a Lie 2-group one can associate a 4d TQFT, and that this TQFT describes a confining phase of a gauge theory with the high-energy gauge group $G$ and the low-energy 't Hooft flux taking values in $\ker\, t$. From now on, we restrict ourselves to this situation. In particular, we will assume that $\dt: \frh\ra \frg$ (the differential of the map $t$) is an isomorphism. 
 
 A 2-connection corresponding to a Lie 2-group is, roughly speaking, a pair $(A,B)$, where $A$ is locally a 1-form with values in the Lie algebra $\frg$ of $G$, and $B$ is locally a 2-form with values in the Lie algebra $\frh$ of $H$. As one goes from chart to chart, these forms transforms as follows:
\begin{eqnarray}
A &\mapsto &gAg^{-1}+gdg^{-1}+\dt(\lambda), \nonumber \\
B &\mapsto & \alpha(g)(B)+d\lambda-\lambda\wedge\lambda \\
 &  &+(\da)(gAg^{-1}+g dg^{-1}+\dt(\lambda))(\lambda) \,. \nonumber
\end{eqnarray}
Here $g$ is a $G$-valued gauge transformation, $\lambda$ is an $\frak h$-valued 1-form, $\dt:{\frak h}\ra\frak g$ is the differential of $t$, and $\da:\frak g\ra {\bf aut} (H )$ is the differential of  $\alpha$. One can check that the combination $F_A-\dt(B)$ transforms as a 2-form in the adjoint representation of $G$.  

The crucial fact used below is that 1-form gauge transformations allow one to shift the 't Hooft flux of $A$ by an arbitrary element of $\pi_1(H)$. Thus t' Hooft flux can be defined only as an element of $\pi_1(G)/\pi_1(H)=\ker\, t$. To demonstrate this, we need to define 2-connections more precisely. Recall that a connection on a principal $G$-bundle over a manifold can be defined by choosing an open cover $\{U_i\}$, $i\in I$ of $M$, so that all charts of the cover and all double, triple, etc. overlaps are contractible, and picking a collections of $\frg$-valued 1-forms $A_i$ on $U_i$ and $G$-valued functions on double overlaps $U_{ij}=U_i\cap U_j$ so that on each $U_{ij}$ one has
$$
A_j=g_{ij} A_i g_{ij}^{-1}+g_{ij} dg_{ij}^{-1},
$$
and on each triple overlap $U_{ijk}=U_i\cap U_j \cap U_k$ one has
\begin{equation}\label{cocycle1}
g_{ik}=g_{jk}g_{ij}.
\end{equation}
Here we implicitly adopted the convention $g_{ij}=g_{ji}^{-1}$ and $g_{ii}=1$ for all $i,j\in I$. 't Hooft flux is then defined as follows. Let $\tG$ be the universal cover of $G$ and ${\tilde t}:\tG\ra G$ be the corresponding homomorphism. For all $U_{ij}$ we pick a lift of $g_{ij}$ to a $\tG$-valued function $\tg_{ij}$ and define on each $U_{ijk}$ the following function:
$$
h_{ijk}=\tg_{ki} \tg_{jk} \tg_{ij}.
$$
Thanks to the equation (\ref{cocycle1}), $h_{ijk}$ takes values in $\ker\, \tilde t=\pi_1(G)$ and defines a Cech 2-cocycle with values in $\pi_1(G)$. One can check that the cohomology class of this cocycle is invariant with respect to gauge transformations of the data $(A_i,g_{ij})$. These gauge transformations are specified by a collection of $G$-valued functions $g_i$ on each $U_i$  and act on $(A_i,g_{ij})$ as follows:
$$
A_i\mapsto g_i A_i g_i^{-1}+g_i dg_i^{-1},\quad g_{ij}\mapsto g_j g_{ij} g_i^{-1}.
$$
Thus we get a well-defined element of $H^2(M,\pi_1(G))$ which is a topologist's version of 't Hooft flux.

The definition of a 2-connection is similar but more complicated \cite{BaezSchreiber,SchreiberWaldorf}. On each chart $U_i$ one needs to specify a $\frg$-valued 1-form $A_i$, on each $U_{ij}$ one needs to specify a $G$-valued function $g_{ij}$ and an $\frh$-valued 1-form $\lambda_{ij}$, and on each $U_{ijk}$ one needs to specify an $H$-valued function $h_{ijk}$ so that the following conditions are satisfied:
\begin{itemize}
\item On each $U_{ij}$ one has $A_j=g_{ij} A_i g_{ij}^{-1}+g_{ij} dg_{ij}^{-1}-\dt(\lambda_{ij})$.
\item On each $U_{ijk}$ one has $g_{ik}=t(h_{ijk})g_{jk} g_{ij}$, and
\begin{eqnarray}
h_{ijk}^{-1}\lambda_{ik} h_{ijk} & = & \da (g_{jk})(\lambda_{ij})+\lambda_{jk} -h_{ijk}^{-1} d h_{ijk} \nonumber \\
 & & -h_{ijk}^{-1}\, \dt^{-1}(A_k) h_{ijk} + \dt^{-1} (A_k) \,. \nonumber
\end{eqnarray}
\item On each $U_{ijkl}$ one has $h_{ijl}h_{jkl} \, = h_{ikl} \cdot \alpha(g_{kl})(h_{ijk}).$
\end{itemize}

Just like in the case of ordinary connections, there is a notion of gauge equivalence of 2-connections. 
Let $(A_i,g_{ij},\lambda_{ij},h_{ijk})$ and $(A_i',g_{ij}',\lambda_{ij}',h_{ijk}')$ be a pair of 2-connections. A gauge-equivalence between them is a $G$-valued  function $g_i$ and an $\frh$-valued 1-form $\lambda_i$ on each $U_i$, together with an $H$-valued function $h_{ij}$ on every $U_{ij}$ such that
\begin{itemize}
\item On each $U_i$ one has 
$$
A_i'=g_i A_i g_i^{-1}+g_i dg_i^{-1}-\dt(\lambda_i).
$$
\item On each $U_{ij}$ one has $g_{ij}'=t(h_{ij})g_j g_{ij} g_i^{-1},$ and
\begin{eqnarray}
\lambda_{ij}' & = & h_{ij} \left(\alpha(g_j)(\lambda_{ij})+\lambda_j\right)h_{ij}^{-1}-\alpha(g'_{ij})(\lambda_i) \nonumber \\
 & & +h_{ij}dh_{ij}^{-1}+h_{ij}\dt^{-1}(A_j') h_{ij}^{-1}-\dt^{-1}(A_j')  \,. \nonumber
\end{eqnarray}
\item On each $U_{ijk}$ one has
$$
h_{ijk}'=h_{ik}\alpha(g_k)(h_{ijk})h_{jk}^{-1}\alpha(g_{jk}')(h_{ij}^{-1}).
$$
\end{itemize}
Note that with our assumptions about $(G,H,t,\alpha)$ one can also write $\alpha(g)(h)=t^{-1}(g)h t^{-1}(g^{-1})$ and $\alpha(g)(\lambda)=t^{-1}(g)\lambda t^{-1}(g^{-1})$ for any $g\in G$, $h\in H$ and $\lambda\in\frh$. These expressions are well-defined, although the map $t^{-1}$ is multivalued.

With this definition of 2-connections and gauge transformations it is not obvious how to define {\it any} analog of the 't Hooft flux invariant under gauge transformations. We will not try to do it here and instead address a closely related question: which 't Hooft line operators can terminate at a point? Here 't Hooft line operators are defined in the usual way, by specifying a singularity in $A$ of a particular kind characterized by an element $m$ of  $\pi_1(G)$. We are going to show that in our TQFT an 't Hooft line operator can terminate if and only if $m$ lies in the subgroup $\pi_1(H)$. Thus 't Hooft flux is conserved only modulo elements of $\pi_1(H)$, as predicted by the monopole condensation picture. 

Consider a 3-sphere whose center is the endpoint of an 't Hooft line. The 't Hooft line operator pierces $S^3$ at a point $p$ which we will call the north pole. The gauge field $A$ is singular there. Everywhere else $A$ is supposed to be nonsingular. Let us cover $S^3$ with two charts so that each chart is homeomorphic to a 3d ball $D^3$ and they overlap over an equatorial region homeomorphic to $S^2\times (-1,1)$. On the chart $U_3$ containing the south pole the field $A$ is nonsingular, and since $U_3$ is contractible, we may assume that $A_3=A\vert_{U_3}$ is a well-defined  $\frg$-valued 1-form. On the other hand, $A$ is singular at the north pole, and after we remove it, the ``northern'' chart becomes topologically nontrivial (homeomorphic to $S^2\times (-1,1))$. Thus $A$ need not be a globally defined 1-form on the ``northern'' chart. In fact, if the 't Hooft flux of the 't Hooft line operator is nontrivial, it {\it cannot} be globally well-defined on the ``northern'' chart. So we have to replace it with two open charts $U_1$ and $U_2$ each of which is homeomorphic to $D^2\times (-1,1)$, so that their overlap $U_{12}$ is homeomorphic to $S^1\times (-1,1)\times (-1,1)$. Accordingly, we have two $\frg$-valued 1-forms $A_1$ and $A_2$ which on $U_{12}$ are related by
$$
A_2=g_{12} A_1 g_{12}^{-1}+g_{12}dg_{12}^{-1}.
$$
The function $g_{12}$ is a $G$-valued function on $U_{12}$. Since $U_{12}$ is homotopically equivalent to $S^1$, $g_{12}$ defines an element of $\pi_1(G)$; this element $m$ is the 't Hooft flux of the 't Hooft line operator. 

Now we would like to construct a well-defined 2-connection on $S^3\backslash p$ which has the properties described in the previous paragraph. If such a 2-connection exists, then the 't Hooft line operator can terminate at a point. 

If $m$ lies in the subgroup $\pi_1(H)$, this can be done as follows. On $U_3$ we let $A_3=0$. We have three double overlaps $U_{12}, U_{13}$ and $U_{23}$. On $U_{12}$ we already have the transition function $g_{12}$, and the consistency condition for the 2-connection on $U_{12}$ requires
$$
\lambda_{12}=0.
$$
On $U_{13}$ we let
$$
g_{13}=1,\quad \lambda_{13}=\dt^{-1}(A_1).
$$
On $U_{23}$ we let
$$
g_{23}=1,\quad \lambda_{23}=A_2.
$$
It is easy to see that, with these choices, the consistency conditions on all double overlaps are satisfied.
There is also one  triple overlap $U_{123}$ on which we need to choose an $H$-valued function $h_{123}$ so that $t(h_{123})=g_{12}^{-1}$ and the following equation holds:
\begin{equation}\label{hgt}
\dt(h_{123}^{-1} dh_{123}^{-1})=g_{12} dg_{12}^{-1}.
\end{equation}
Since we assumed that the loop in $G$ defined by $g_{12}$ is in the image of the map $t:\pi_1(H)\ra\pi_1(G)$, there is no obstruction to finding an $H$-valued function $h_{123}$ such that $t(h_{123})=g_{12}$. This $h_{123}$ solves the above equation, so we are done.

Conversely, suppose we managed to find a consistent 2-connection on $S^3\backslash p$ with the desired properties. In particular, this implies that on $U_{123}$ the following equation is satisfied:
$$
g_{13}g_{21}g_{32}=t(h_{123}).
$$
This equation implies that the loop in $G$ defined by the function $g_{13}g_{21}g_{32}: U_{123}\ra G$ lifts to a loop in $H$, and therefore the corresponding element in $\pi_1(G)$ lies in the subgroup $\pi_1(H)$. But this element in $\pi_1(G)$ coincides with the 't Hooft flux of the 't Hooft line operator. Indeed, both $U_{13}$ and $U_{23}$ are contractible (in fact, homeomorphic to $D^2\times (-1,1)$), so both $g_{13}$ and $g_{32}$ are homotopic to the constant map. Hence $g_{13}g_{21}g_{32}$ is homotopic to $g_{21}$, and defines the same element in $\pi_1(G)$ as $g_{21}$. 

Thus we have shown that an 't Hooft line can terminate at a point if and only if its 't Hooft flux is in the subgroup $\pi_1(H)$. 

\subsubsection{A TQFT for a nonabelian confining phase}

We have seen that the kinematic structure of nonabelian 2-connections makes them a suitable candidate for field variables describing a confining phase of a nonabelian gauge theory. It remains to write down an action for such a theory.  By analogy with the abelian case, consider the following action:
\begin{equation}\label{nonabTQFT}
S_{top}=\int {\rm Tr} ((F_A-\dt(B))\wedge b),
\end{equation}
where $b$ is a 2-form with values in the adjoint representation of $G$. This action is obviously gauge-invariant and diffeomorphism-invariant. We propose it as the action describing the confining phase of a gauge theory with gauge group $G$ and monopole condensate with magnetic charges in the subgroup $\pi_1(H)$ of $\pi_1(G)$. 

In the abelian case this TQFT agrees with the previous discussion. For example, suppose we take $G=H=U(1)$, take $t$ to be a degree $n$ covering map, $\dt=n$, and take $\alpha$ to be the constant map which sends the whole $G$ to the identity element of $\Aut (H)$. Then gauge transformations reduce to
\begin{equation}\label{abelian1form}
A\mapsto A+i d\log g+n\lambda,\quad B\mapsto B+d\lambda,
\end{equation}
where $g$ is a $U(1)$ gauge transformation, and the action becomes
$$
S_{top}=\int (F_A-nB)\wedge b.
$$

In general, the action (\ref{nonabTQFT}) describes a theory which has no local degrees of freedom. Indeed, the equations of motion set $F_A=\dt(B)$, and since $\dt$ is invertible, locally $B$ is expressed through $F_A$, and $A$ itself can be locally eliminated using a 1-form gauge transformation. Nonlocal observables are 't Hooft loops, whose flux takes values in $\pi_1(G)/\pi_1(H)={\rm ker}\, t$, and electric surface operators which measure the flux of $B$ through an oriented surface. We have seen above how to define such a gauge-invariant flux for a spherical surface only, but it is possible to define it for surfaces of arbitrary genus \cite{SchreiberWaldorf}. Electric surface operators are labeled by elements of $\Hom({\rm ker}\, t,U(1))$.  Wilson loops cannot be defined at all because they are not invariant under 1-form gauge transformations; this reflects confinement in the underlying non topological gauge theory. 

If $G=H$ and $t$ is the identity map, the TQFT (\ref{nonabTQFT}) is completely trivial since it does not admit any nontrivial observables. In general, the structure of observables suggests that it is equivalent to the topological gauge theory with a discrete gauge group $\Hom({\rm ker}\, t,U(1))$. Note that this group is always abelian.   R. Thorngren \cite{Ryan} computed the partition function of a 4-manifold, the states space of a 3-manifold,  and the category of a 2-manifold for the TQFT (\ref{nonabTQFT}) and verified that they agree with the corresponding objects  for the discrete gauge theory. 

\subsection{A TQFT for a nonabelian oblique confining phase}

The action (\ref{nonabTQFT}) can be modified by a theta-term for the gauge field:
$$
S^\theta_{top}=\int {\rm Tr} ((F_A-\dt(B))\wedge b)+\frac{i\theta}{8\pi^2} \int {\rm Tr} (F_A\wedge F_A).
$$
 We proposes that this TQFT describes a phase with a dyonic condensate. This is reflected in the fact that 't Hooft loop operators are no longer invariant under 1-form gauge transformations thanks to the last term in the action. This is interpreted as confinement of 't Hooft loops . Wilson loops are also not gauge-invariant by themselves. However, it is possible  to construct Wilson-'t Hooft loop operators which are invariant under all gauge transformations. The flux of such Wilson-'t Hooft operators is measured by electric surface operators which are defined in the same way as before.
 
 The theta-angle angle has to be quantized in order for $\exp(-S^\theta_{top})$ to be invariant under 1-form gauge transformations. Indeed, consider a connection $A$ which is equivalent to zero thanks to a 1-form gauge invariance. Such a connection is obtained by embedding an $H$-instanton into $G$ using the homomorphism $t$. Requiring $\exp(-S^\theta_{top})=1$ puts a quantization condition on $\theta$. For concreteness, consider the case when $G=SU(N)/\ZZ_N$, $H=SU(N)/\ZZ_p$, where $p$ divides $N$, with $t$ being the degree $N/p$ cover.  On a general manifold the instanton number for an $SU(N)/\ZZ_p$ gauge field is an integer multiple of $1/p$, and therefore the quantization condition says $\theta=2\pi p n$, where $n\in\ZZ$. This does not mean that $\theta$ is trivial, since for $G=SU(N)/\ZZ_N$ the periodicity of $\theta$ is $2\pi N$ rather than $2\pi$ (see \cite{AST} for a recent discussion). Thus $\theta$ can have $N/p$ physically distinct values. One possible interpretation is that if monopoles with 't Hooft flux $p$ in a physical $SU(N)/\ZZ_N$ have condensed, the theta-angle must flow in the IR to one of these quantized values.
 
\section{Discussion}

\subsection{Higgs vs. confinement}

Note that given a central extension of a compact Lie group $G$ by a finite abelian group $\Gamma$,
\begin{equation}\label{extension1}
1\ra \Gamma\ra H\ra G\ra 1,
\end{equation}
we can construct two TQFTs: a confining one with the UV gauge group $G$ and 't Hooft fluxes taking values in $\Gamma$, and a Higgsed one with high energy gauge group $H$ and low-energy gauge group $\Gamma$. 't Hooft loops in the former TQFT are labeled by elements of $\Gamma$, while Wilson loops in the latter TQFT are labeled by elements of $\hat\Gamma={\rm Hom}(\Gamma,U(1))$ (the Pontryagin dual of $\Gamma$). If we consider Langlands-dual groups $\hat H$ and $\hat G$, there will be a homomorphism in the opposite direction, $\hat t: \hat G\ra\hat H$ and a central extension
\begin{equation}\label{extension2}
1\ra \hat\Gamma\ra \hat G\ra \hat H\ra 1
\end{equation}
There will again be a pair of TQFTs, with 't Hooft loops in the confining TQFT labeled by elements of $\hat\Gamma$ and Wilson loops in the Higgsed TQFT labeled by elements of $\Gamma$. Presumably the confining TQFT corresponding to the extension (\ref{extension1}) is isomorphic to the Higgs TQFT corresponding to the extension (\ref{extension2}), and vice versa. This can be thought of as a topological version of electric-magnetic duality. 

Note that there are also more general ``Higgs'' TQFTs constructed from finite subgroups $\Gamma$ which are not central or even normal. It appears that such more general TQFTs do not have a confining counterpart.  This suggests that gauge theories whose gauge group is Higgsed down to a subgroup which is not central do not admit a dual confining description.

\subsection{Lattice gauge theory}

There is ample evidence from lattice gauge theory that pure Yang-Mills theory with gauge group $G=SU(N)$ is confining in the sense of the Wilson criterion. That is, Wilson loops in the representations which transform nontrivially under the center of $G$ obey the area law.  On the other hand, if we take the gauge group to be $SU(N)/\ZZ_N$, then all allowed Wilson loops obey the perimeter law, and so do 't Hooft loops.   Both $G=SU(N)$ and $G=SU(N)/\ZZ_N$ theories have a mass gap, and one may ask which TQFT describes their low-energy limit. 

In the $SU(N)$ theory the TQFT is obviously trivial, because (1) no topologically nontrivial 't Hooft loops can be defined, and (2) all Wilson loops are either confined or screened: the ones which transform nontrivially under the center of $SU(N)$ are confined, while the rest are screened by gauge bosons and have no long distance effects. In the $SU(N)/\ZZ_N$ case all Wilson loops are screened by gauge bosons, but the situation with 't Hooft loops depends on the charges of monopole condensate. TQFT is precisely sensitive to the subgroup of $\pi_1(SU(N)/\ZZ_N)=\ZZ_N$ which is generated by the charges of the monopole condensate. For $N$ prime there are only two possibilities (either all possible monopoles have condensed, or only monopoles with a trivial 't Hooft flux have condensed), but in general possible phases correspond to divisors of $N$.

It is natural to ask how to realize all these phases of $SU(N)/\ZZ_N$ Yang-Mills theory using lattice gauge theory. In fact the answer to this question has been given a long time ago \cite{MP}. The idea is to use a Villain-type formulation of $SU(N)/\ZZ_N$ gauge theory, where the link variables $U_l$ take values in $SU(N)$, but there are also plaquette variables $w_P$ taking values in $\ZZ_N\subset U(1)$. The role of the plaquette variables is to ensure the invariance of the action with respect to the discrete 1-form gauge transformation 
\be\label{discretegauge}
U_l\mapsto \eta_l U_l,\quad \eta^N=1,\quad w_P\mapsto w_P \prod_{l\in\partial P} \eta_l.
\ee
The plaquette variables can be thought of as describing lattice monopoles with $\ZZ_N$-valued 't Hooft flux. More precisely, the magnetic flux through a cube $c$ is given by the product 
$$
m_c=\prod_{P\in \partial c} w_P,
$$
so monopole worldlines should be thought of as living on the links of the dual lattice. 

In the usual Villain model one simply sums over all possible values of the variables $w_P$. This lattice model corresponds to a phase where monopoles with all possible 't Hooft fluxes have condensed, and accordingly the low-energy phase is described by a nonabelian confining TQFT associated to the trivial cover $SU(N)/\ZZ_N\ra SU(N)/\ZZ_N$. However, it is also possible to put a constraint on the magnetic flux of lattice monopoles. This constraint must respect the gauge-invariance (\ref{discretegauge}). A natural gauge-invariant constraint is
$$
m_c=1,\quad \forall c.
$$
This ensures the absence of lattice monopoles with a nontrivial 't Hooft flux. (The 't Hooft flux through a nontrivial homology 2-cycle is still allowed to be nontrivial). This lattice model corresponds to a nonabelian confining TQFT associated to the universal cover $SU(N)\ra SU(N)/\ZZ_N$.  If $N$ is not a prime number, one can also pick its divisor $p$ and consider a weaker constraint
\begin{equation}\label{cube}
m_c^p=1,\quad \forall c.
\end{equation}
This lattice model corresponds to a phase where monopoles with 't Hooft fluxes divisible by $p$ have condensed; this phase is described in the continuum limit by a nonabelian gerbe TQFT associated to the cover $SU(N)/\ZZ_p\ra SU(N)/\ZZ_N$.

One can think of this lattice model as the result of coupling an abelian TQFT for a $\ZZ_N$-valued B-field described by plaquette variables $w_P$ to a lattice $SU(N)$ gauge theory. This coupling confines the center of $SU(N)$ already at the microscopic level and therefore gives $SU(N)/\ZZ_N$ gauge theory in the continuum. For $p=1$ this coupling can be described by a crossed module $t:\ZZ_N\ra SU(N)$, where $t$ is the obvious embedding,  and the action of $SU(N)$ on $\ZZ_N$ is the trivial one. The main difference compared to the case considered in the bulk of the paper is that the quotient $SU(N)/\ZZ_N$ is not finite, and accordingly the action is not topological but rather is a lattice version of
$$
\int {\rm Tr} ||F-t(B)||^2. 
$$
For $p>1$ the TQFT is a bit more complicated because the constraint on cubes (\ref{cube}) can be most naturally interpreted in terms of a $\ZZ_p$-valued variable living on cubes, i.e. a discrete 3-form.

\section*{Acknowledgements}

We are grateful to Nathan Seiberg for a collaboration during various stages of this project and to Gregory Moore for a discussion.  This work was supported in part by the DOE grant DE-FG02-92ER40701 and 
by the National Science Foundation under Grant No. PHYS-1066293 and the hospitality of the Aspen Center for Physics.

%%%%%%%%%%%%%%%%%%%%%%%%%%%%%%%%%%%%%%%%%%%%%%%%%%%%%%%%%%%%%%%%%%

\end{document}